\documentstyle[psfig]{elsart}
\begin{document}
\begin{frontmatter}
\vspace*{-.25in}
\title{$\Lambda$ hyperons in 2 $A\,$GeV Ni + Cu collisions} 
\vspace*{-.25in}
\collab{EOS Collaboration}
\vspace*{-.15in}
\author[Kent]{M.~Justice\thanksref{MJ}},
\author[Catania]{S.~Albergo},
\author[LBNL]{F.~Bieser},
\author[Davis]{F.P.~Brady},
\author[Catania]{Z.~Caccia},
\author[Davis]{D.A.~Cebra},
\author[TAMU]{A.D.~Chacon},
\author[Davis]{J.L.~Chance},
\author[Purdue]{Y.~Choi\thanksref{YChoi}},
\author[Catania]{S.~Costa},
\author[Purdue]{J.B.~Elliott}, 
\author[Purdue]{M.L.~Gilkes},
\author[Purdue]{J.A.~Hauger}, 
\author[Purdue]{A.S.~Hirsch}, 
\author[Purdue]{E.L.~Hjort},
\author[Catania]{A.~Insolia},
\author[Kent]{D.~Keane},
\author[Davis]{J.C.~Kintner\thanksref{JKintner}},
\author[LBNL]{M.A.~Lisa\thanksref{MLisa}}, 
\author[Kent]{H. Liu},
\author[LBNL]{H.S.~Matis},
\author[SUNYSB]{R.~McGrath},
\author[LBNL]{M.~McMahan},
\author[LBNL]{C.~McParland},
\author[LBNL]{D.L.~Olson},
\author[Davis]{M.D.~Partlan}, 
\author[Purdue]{N.T.~Porile}, 
\author[Catania]{R.~Potenza}
\author[LBNL]{G.~Rai},
\author[LBNL]{J.O.~Rasmussen},
\author[LBNL]{H.G.~Ritter},
\author[Catania]{J.~Romanski\thanksref{JRomanski}}, 
\author[Davis]{J.L.~Romero},
\author[Catania]{G.V.~Russo},
\author[Purdue]{R.P.~Scharenberg},
\author[Kent]{A.~Scott\thanksref{AScott}}, 
\author[Kent]{Y.~Shao\thanksref{YShao}},
\author[Purdue]{B.K.~Srivastava}, 
\author[LBNL]{T.J.M.~Symons},
\author[Purdue]{M.~Tincknell},
\author[Catania]{C.~Tuv\`{e}},
\author[Kent]{S.~Wang},
\author[Purdue]{P.G.~Warren}, 
\author[Kent]{D.~Weerasundara\thanksref{Dhammika}},
\author[LBNL]{H.H.~Wieman},
\author[LBNL]{T.~Wienold\thanksref{TWienold}}
\author[TAMU]{K.~Wolf}

\address[Kent]{Kent State University, Kent, OH 44242}
\address[Davis]{University of California, Davis, CA 95616}
\address[Catania]{Universit\'{a} di Catania and Instituto Nazionale di Fisica
Nucleare-Sezione di Catania
95129 Catania, Italy}
\address[LBNL]{Nuclear Science Division, Lawrence Berkeley National
Laboratory, Berkeley, CA 94720}
\address[Purdue]{Purdue University, West Lafayette, IN 47907}
\address[SUNYSB]{State University of New York at Stony Brook, Stony Brook, New
York 11794}
\address[TAMU]{Texas A\&M University, College Station, TX  77843}
\date{August 3, 1997}
\thanks[MJ]{ Present address: 
Brookhaven National Laboratory, Upton, NY 11973.}
\thanks[YChoi]{ Present address: 
Sung Kwun Kwan University, Suwon, Rep. of Korea, 440-746.}
\thanks[JKintner]{ Present address:
St. Mary's College of California, Moraga, CA 94575.}
\thanks[MLisa]{ Present address:
Ohio State University, Columbus, OH 43210.}
\thanks[JRomanski]{ Present address: The Svedberg Laboratory, University of Uppsala, S751-21 Uppsala.}
\thanks[AScott]{ Present address: University of Wisconsin-Stout, Menomonie, WI 54751.}
\thanks[YShao]{ Present address: 
Crump Institute for Biological Imaging, UCLA, Los Angeles, CA 91776.}
\thanks[Dhammika]{ Present address: 
University of Washington, Seattle, WA 98195.}
\thanks[TWienold]{ Present address:
Universit\"{a}t Heidelberg, Germany}

\begin{abstract}
A sample of $\Lambda$'s produced in 
2 $A\,$GeV $^{58}$Ni + $^{\mbox{nat}}$Cu
collisions has been obtained with the 
EOS Time Projection Chamber at the Bevalac.
Low background in the invariant 
mass distribution allows for the 
unambiguous demonstration of
$\Lambda$ directed flow.
The $\Lambda$ $m_{T}$ spectrum at mid-rapidity
has the characteristic shoulder-arm shape
of particles undergoing radial
transverse expansion.
A linear dependence of $\Lambda$ multiplicity on impact parameter
is observed, from which a total 
$\Lambda + \Sigma^{0}$ production cross section of $112 \pm 24$ mb 
is deduced. Detailed comparisons with the ARC and RVUU models
are made.
\end{abstract}
\vspace*{-.15in}
\end{frontmatter}

\section{Introduction}

Relativistic nucleus-nucleus collisions 
provide a unique opportunity for studying
hot dense hadronic matter.
At beam energies below 2 $A\,$GeV
the focus has been on extracting information about
the nuclear matter equation of state
from measurements of the
collective flow of  nucleons and light
fragments \cite{hgr97}. 
As the beam energy is increased, however,
a complete description of
the EoS should also include 
strange particle degrees of freedom.
Recent theoretical studies 
have indicated that the in-medium properties of 
strange particles in nuclear matter at high densities  
and temperatures may be constrained by
accurate measurements of their yields, spectra, 
and flow in heavy-ion collisions \cite{li95,li96}. 
Preliminary experimental evidence of a
directed flow signal for lambda particles has been provided
by the EOS collaboration \cite{qm95} and 
by the FOPI collaboration \cite{fopi95}. In this Letter we
present the completed analysis of $\Lambda$ sideward flow
in the EOS data. In addition we present for the first time 
the EOS results on the total yield,
the yields as a function of
centrality and rapidity, and the
collective transverse expansion of $\Lambda$'s.

\section{Experiment and data analysis}

A full description of the EOS experimental setup can
be found in Ref.s~\cite{rai90,wie91,hjort93}.
The heart of the setup is a
Time Projection Chamber (TPC) 
which provides continuous three dimensional tracking and 
particle identification  for particles  with $Z \leq 8$.
The EOS TPC has a rectangular geometry and operated
in anti-parallel 1.3 T $\vec{\mathrm{B}}$ and 
120 V/cm $\vec{\mathrm{E}}$ fields for this experiment.
The target was situated as close to the
TPC as possible, approximately 15 cm 
upstream of the first TPC pad row, in order
to maximize acceptance. Because 
this placed it in a high magnetic field,
copper rather than nickel was used
as the target material. The 
$^{58}$Ni beam supplied by the Berkeley
Bevalac had an energy of 1.97 $A\,$GeV
at the center of the target. 
The hardware trigger consisted of beam 
defining counters located some distance
upstream and a trigger counter located just 
downstream of the target.
The discriminator threshold on the
downstream counter was set to
veto the $\sim$25$\%$ most peripheral
events. This trigger cutoff produces 
negligible bias in the data sample  
since, as demonstrated below, 
$\Lambda$'s are predominantly 
produced in central events.

\begin{figure}
\centerline{\psfig{figure=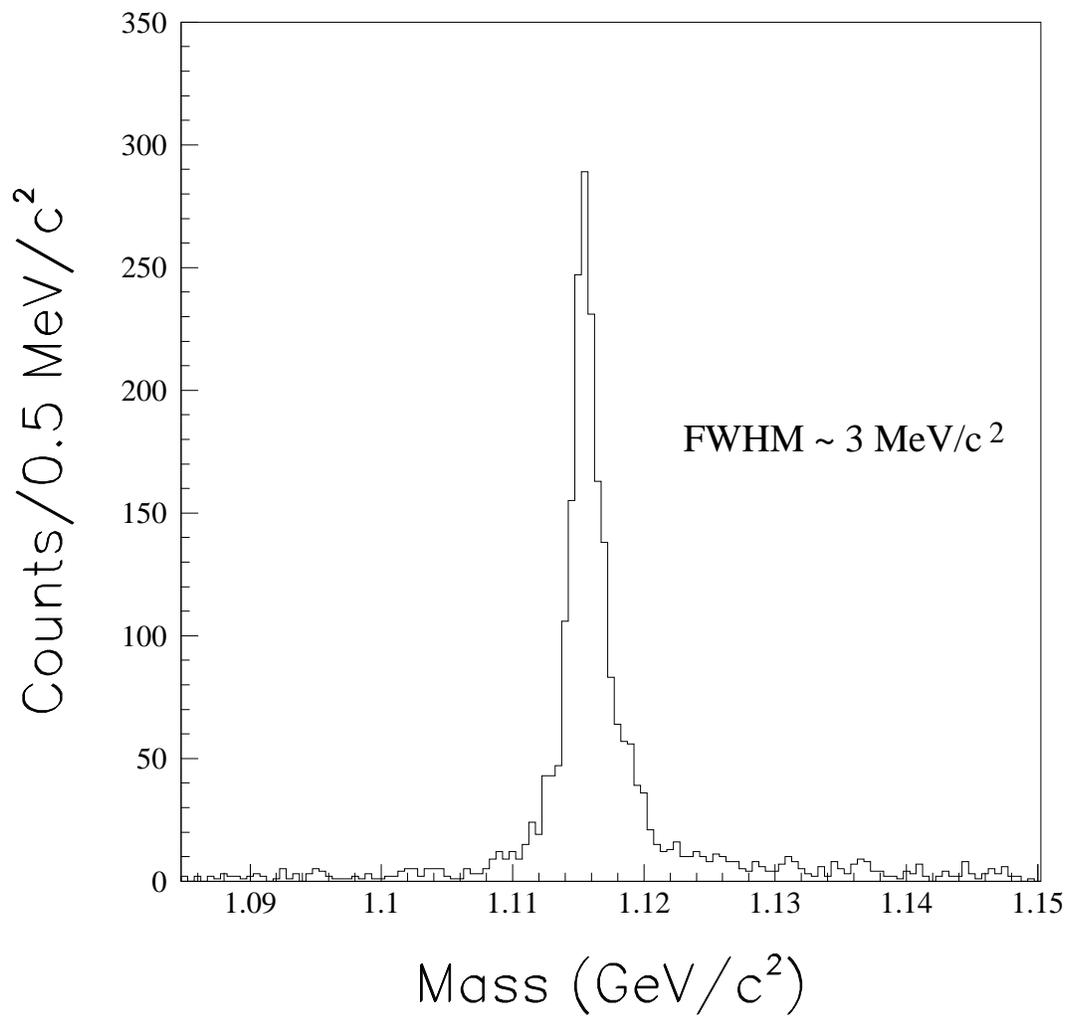,height=6in}}
 \caption { $\Lambda$ invariant mass spectrum.}
 \label{MLAM}
\end{figure}

$\Lambda$'s are reconstructed through the charged
particle decay: \mbox{$\Lambda \longrightarrow p + \pi^{-}$}, which
has a branching ratio of approximately 64$\%$. After all
TPC tracks in an event are found and the overall
event vertex has been determined,
each pair of $p\pi^{-}$ tracks
is looped over and their point of
closest approach is calculated. Pairs
whose trajectories intersect at a point other
than the main vertex are fit with  a
V0 hypothesis from which an invariant mass 
and momentum are extracted. The $\Lambda$ candidates
are then passed through a neural network filter \cite{nnet96}
in order to reduce the combinatoric background. 
The resulting invariant mass distribution is 
shown in Fig.~\ref{MLAM}. Cutting around the peak 
in the distribution (1112 MeV/$c^{2}$ $\leq M_{\Lambda} \leq$ 1120
MeV/$c^{2}$) results in a sample of 1797  
$\Lambda$ candidates, of which around 40
are estimated to be background.

Acceptance correction factors have been calculated
by passing a sample of \mbox{$7.2 \times 10^{5}$}
minimum bias 2 $A\,$GeV Ni + Cu ARC 
events through a detailed GEANT simulation of the EOS
detector system and then through the same analysis
chain as actual data. The ARC model \cite{arc92} has been successful
at reproducing a wide range of inclusive observables
at AGS energies and it has been used to
study directed flow in Au + Au collisions at Bevalac
energies \cite{kah94}. The parameterization of the 
TPC response in the EOS GEANT simulation code is
similar to the parameterization in
the FST simulation program used by the STAR collaboration \cite{star92}.
Hit merging, spatial resolution, and ADC resolution
parameters were empirically adjusted to match
the data. The overall width and background
of the reconstructed Monte Carlo $\Lambda$'s 
are in good agreement with the data of
Fig.~\ref{MLAM}. 

Azimuthally integrated $\Lambda$ efficiency factors are calculated 
on a $15 \times 25$ grid in $(y, m_{T})$ space from the formula:
\begin{equation}
\epsilon(y_{i},m_{T,j})  =  \frac{N_{rec}(y_{i},m_{T,j})}{N_{MC}(y_{i},m_{T,j})} \,\,\, ,
\end{equation}
where $N_{rec}$ and $N_{MC}$ are the reconstructed
and input number of $\Lambda$'s, respectively, in the given bin.
In principle the efficiency factors also depend on
the multiplicity or impact parameter. 
Generating a sufficient number of GEANT events
to calculate three dimensional efficiencies is
impractical in the present case, however.  
Therefore, the approximation has been
made that the efficiency factorizes into two pieces:
$\epsilon(b,y,m_{T}) = \epsilon(b) \cdot \epsilon(y,m_{T})$;
with the $\epsilon(b)$'s being given by:   
\begin{equation}
\epsilon(b_{i})  =  \frac{N_{rec}(b_{i})}{N_{MC}(b_{i})} \,\,\, ,
\end{equation}
averaged over all values of $y$ and $m_{T}$.
In central and semi-central events tracking 
efficiencies are independent of impact parameter
for the relatively light \mbox{Ni + Cu} system
(maximum TPC multiplicity of around 70 tracks).
In peripheral collisions,
however, the presence of a forward going high
Z fragment creates a hole in the TPC
acceptance near beam rapidity.  Although the
factorization assumption is not valid
for such events, the overall systematic error
introduced is expected to be small, as the
hole occurs in a region where there are 
few $\Lambda$'s.  
At mid-rapidity, where the majority of $\Lambda$'s are
produced, the efficiency factors are very flat
versus $p_{T}$ over the entire range of impact parameters.

The overall acceptance for 
\mbox{$\Lambda \longrightarrow p + \pi^{-}$},
averaged over $b$, $y$, and $p_{T}$,
is calculated to be about 16$\%$. 
Approximately half of the 
loss is due to geometrical acceptance.
Determination of the precise fraction
decaying outside the solid angle 
of the detector is facilitated by the 
fact that, to first order, ARC 
reproduces the shape of the 
experimental distribution
reasonably well. To second order, however,
a small systematic uncertainty
of $<10\%$ might be introduced
by the extrapolation of efficiency
factors to regions of phase space
where there are no counts.
Some $\Lambda$'s are also lost
through inefficiencies in the
track reconstrucion software.
According to the simulations, 
over ninety percent of the 
$\Lambda$'s which decay within 
the geometrical acceptance are
reconstructed. Event display
comparisons of data with Monte Carlo  
suggest that the true tracking
efficiency may be lower
by a few percent.
The remainder of the experimental
inefficiency is due to  
the cuts used to
reduce the combinatoric background.
More than $50\%$ of the true 
$\Lambda$'s which are  reconstructed 
get thrown out by these cuts.
Systematic error in the 
estimation of the magnitude
of this loss mechanism is believed
to be small. 
The total systematic error 
in the final cross section from
all contributions is estimated
to be ten to fifteen percent --- 
half the size of the statistical error.   

The open channels for $\Lambda$ production in nucleon-nucleon
collisions at this beam
energy are $NN \longrightarrow \Lambda KN$ (1.58 GeV),
$NN \longrightarrow \Sigma^{0} KN$ followed
by $\Sigma^{0} \longrightarrow \Lambda\gamma$ (1.79 GeV), and
$NN \longrightarrow \Lambda KN\pi$ (1.96 GeV).
$\Lambda$'s originating from $\Sigma^{0}$ decay 
($c\tau = 2.2\times 10^{-9}$cm) are experimentally
indistinguishable from primary $\Lambda$'s.
In the ARC event sample approximately $23\%$
of the $\Lambda$'s arise from $\Sigma^{0}$ decay.
These $\Lambda$'s are included in all 
of the ARC comparisons shown below. 
All spectra, except for the sideward flow
plots of Fig.~\ref{FLOW}, have also 
been corrected for branching ratios.
The dependence of $\Lambda$ yield 
upon centrality produces a natural 
weighting towards central events  
but no explicit centrality cuts 
have been applied.
Attempts to study the effects of centrality cuts on the 
flow analyses were inconclusive 
due to insufficient statistics.

\section{Results}

\begin{figure}
\centerline{\psfig{figure=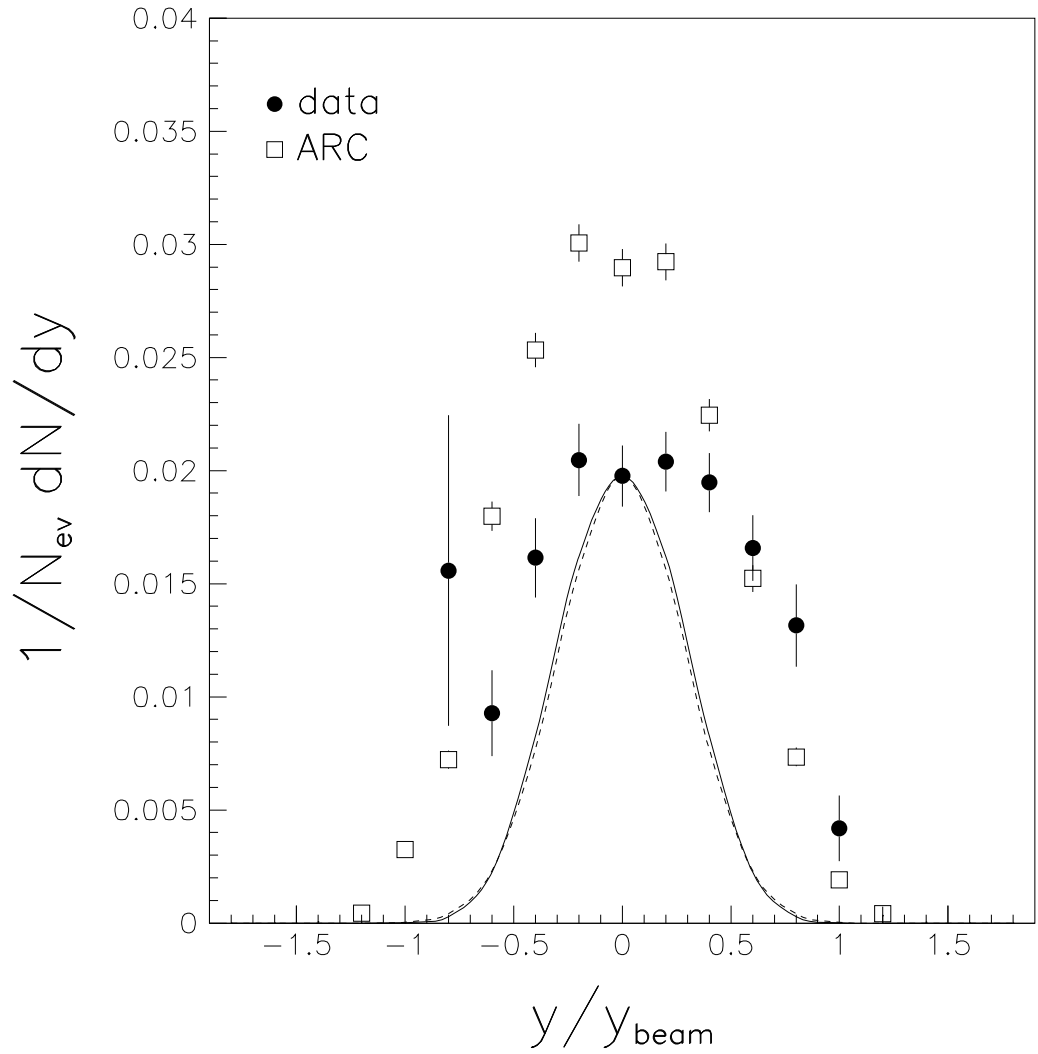,height=6in}}
 \caption { $\Lambda$ rapidity spectra.}
 \label{YLAM}
\end{figure}

The efficiency-corrected $\Lambda$ center-of-mass 
rapidity  spectrum, normalized to
beam rapidity, is shown in Fig.~\ref{YLAM} along with
the ARC comparison. Both spectra
peak near $y=0$ as expected for
this nearly mass-symmetric system. Within statistics,
the widths of the distributions are similar; however,
ARC appears to overestimate the total yield by 
approximately $50\%$. This is not necessarily
a deficiency of the cascade assumption, however,
as there is a relatively large uncertainty in the 
$\Lambda$ cross section 
in $pp$ collisions below 2 GeV.
The solid curve of Fig.~\ref{YLAM}
represents the rapidity spectrum of an 
isotropic thermal distribution
at the temperature of 106 MeV which is
obtained by fitting
an exponential to the mid-rapidity transverse mass spectrum
of the data. The rapidity spectra for both the data and the 
model are  considerably broader.
This  most likely indicates
an insufficient number
of rescatterings for the 
$\Lambda$'s --- which have 
a well known forward-backward 
distribution in elementary $pp$ reactions ---
to become  completely isotropic.

\begin{figure}
\centerline{\psfig{figure=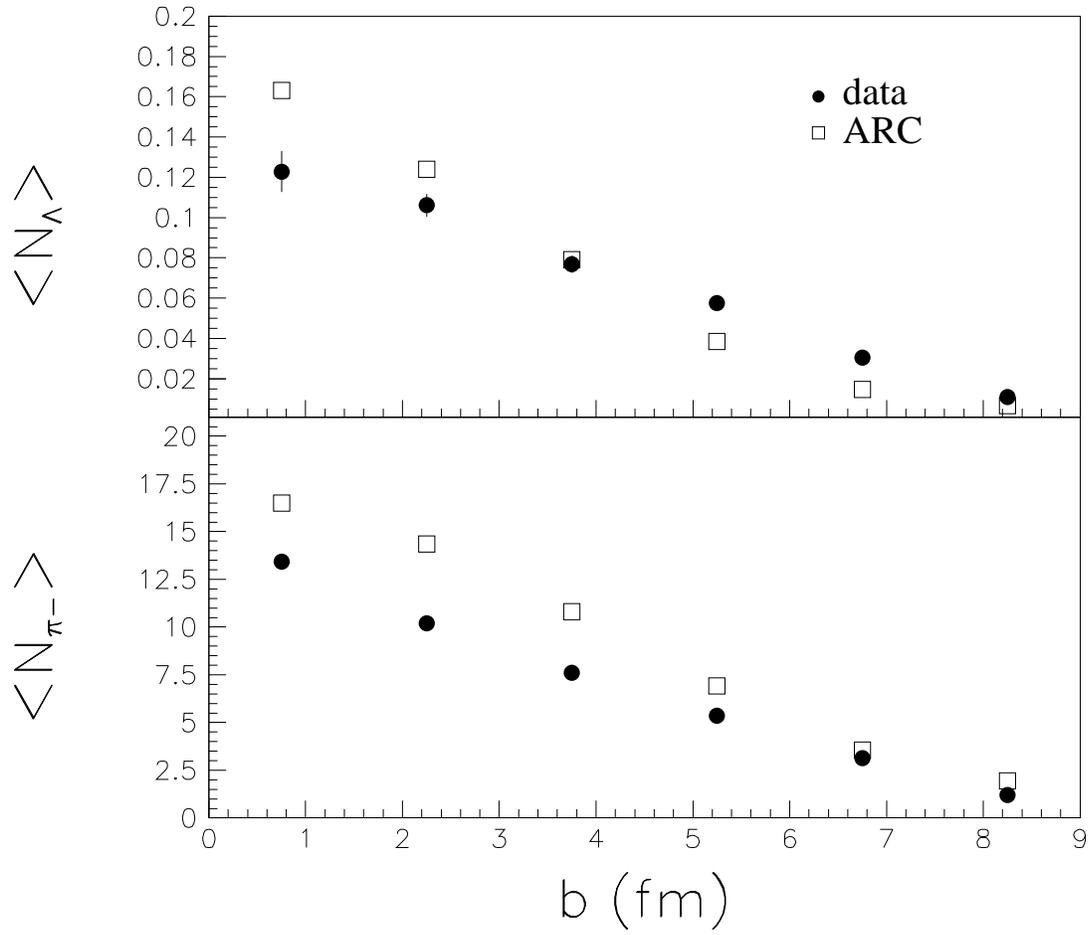,height=6in}}
 \caption { Mean $\Lambda$ and $\pi^{-}$
multiplicities  versus impact parameter.}
 \label{YIELD}
\end{figure}

The dependence of 
$\Lambda$ yield upon event centrality
is shown in the top panel of Fig.~\ref{YIELD}. The
impact parameter scale in this figure was derived
by assuming a monotonic dependence of total observed
charged particle multiplicity in the TPC versus
impact parameter \cite{cav90}. 
The data are well 
described by a straight line fit of $\langle N_{\Lambda} \rangle$
versus $b$: $\langle N_{\Lambda} \rangle = 0.1361(45) -
0.0153(7)b$ with a $\chi^{2}/\nu$ of 0.6.
The ARC model, on the other hand, predicts 
a linear dependence of $\Lambda$ yield 
on the number of participants rather than
on impact parameter.
When plotted versus the number of
participants, the data
bend over and flatten out 
for the most central collisions.
This behavior could be suggestive of 
a kind of ``shadowing'' effect in
which late arriving nucleons from 
the projectile and target suffer
softer collisions with matter 
already stopped in the collision zone.
Since the initial beam energy is 
only marginally above threshold to 
begin with, $\Lambda$ production
from these late primary nucleons
might be suppressed. The fact that
the effect doesn't also show up in 
the cascade model, however, tends to
cast doubt on this explanation. 

The efficiency-corrected mean $\pi^{-}$ 
multiplicities  versus  impact parameter are shown 
in the bottom panel of Fig.~\ref{YIELD}.
The $\pi^{-}$ efficiency 
is approximately $60\%$ and independent of $b$.
The $\pi^{-}$ multiplicities of the data
are not as well fit  by a straight
line as are the $\Lambda$'s. 
The $\langle \pi^{-} \rangle$ 
dependence upon participant number
is reasonably linear, however, consistent
with  previous observations \cite{har85}.    
Within statisitics, the $\Lambda/\pi^{-}$ ratio 
is constant as a function of  $b$ with
a value of $0.010 \pm 0.001$.
For the most central 
events, ARC overpredicts the $\pi^{-}$
yield by about $20\%$.
ARC pion multiplicities bend over significantly
when plotted versus the number of participants.
The net result is a strong dependence of
the $\Lambda/\pi^{-}$ ratio on 
centrality in ARC.

The straight line fit of the
upper panel of Fig.~\ref{YIELD}
intersects the abscissa at $b=8.9\pm0.7$ fm.
Integrating the 
fit out to this value of $b$
yields a total cross section 
of \mbox{$\sigma_{\Lambda + \Sigma^{0}} = 112 \pm 24$ mb.}
As mentioned above, a
systematic error 
of the order of $15\%$
from uncertainties in
the acceptance corrections 
should be attached to this result.
For the central 180 mb of the total reaction 
cross section, corresponding to impact parameters
$b < 2.4$ fm,  the extracted cross section is 
\mbox{$\sigma_{\Lambda + \Sigma^{0}} = 20\pm 3$ mb.}
This is somewhat higher than, but not
in disagreement with,  the $7.6\pm 2.2$ mb
cross section for central 
1.8 $A\,$GeV Ar + KCl collisions of 
Ref.~\cite{harris81} when scaled by 
the approximately $50\%$ increase in the
number of participants.  
The parameterization used in Ref.~\cite{li96}
for the $\Lambda$ production cross section
predicts only a $10\%$ increase in yield
in going from a beam energy of \mbox{1.8 $A\,$GeV} 
to \mbox{2.0 $A\,$GeV.}

\begin{figure}
\centerline{\psfig{figure=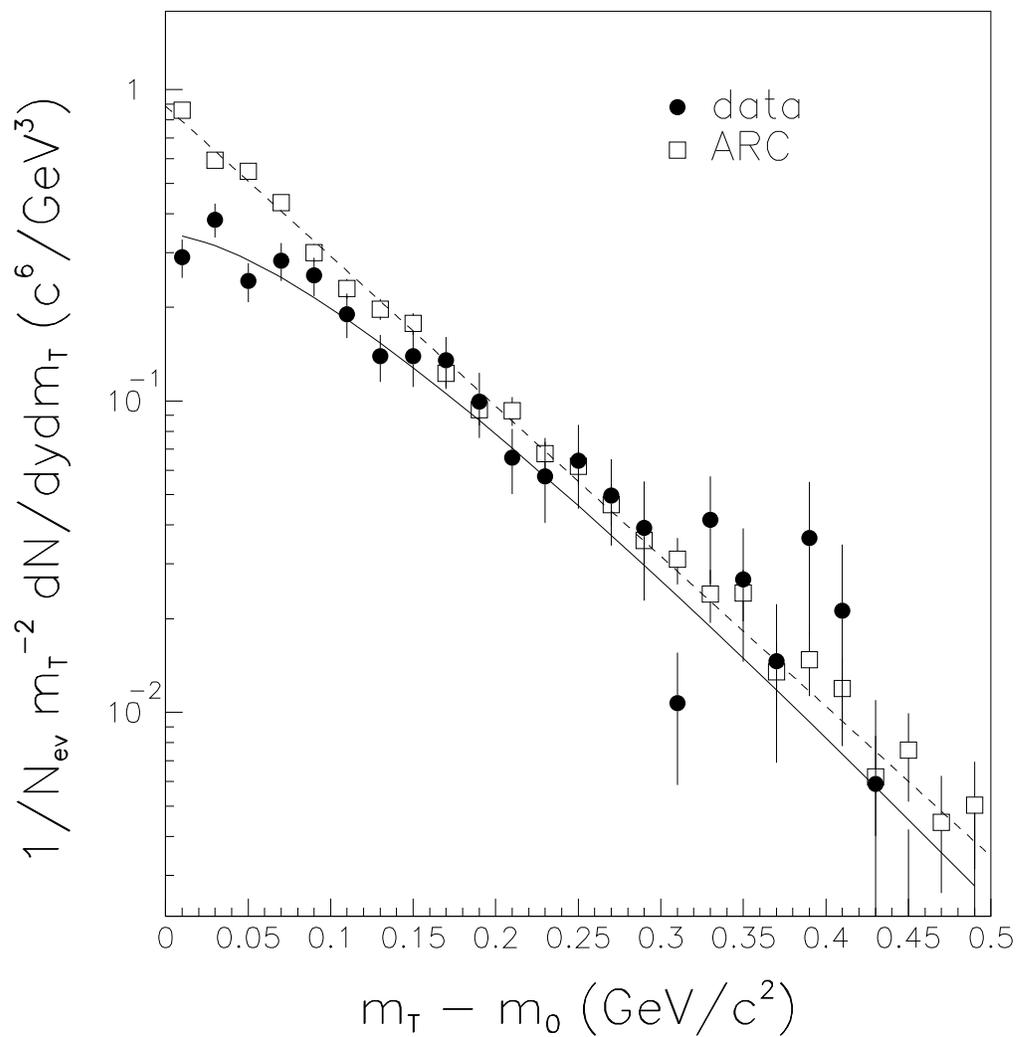,height=6in}}
 \caption { Mid-rapidity $\Lambda$ transverse mass spectra.}
 \label{MTLAM}
\end{figure}

\begin{figure}
\centerline{\psfig{figure=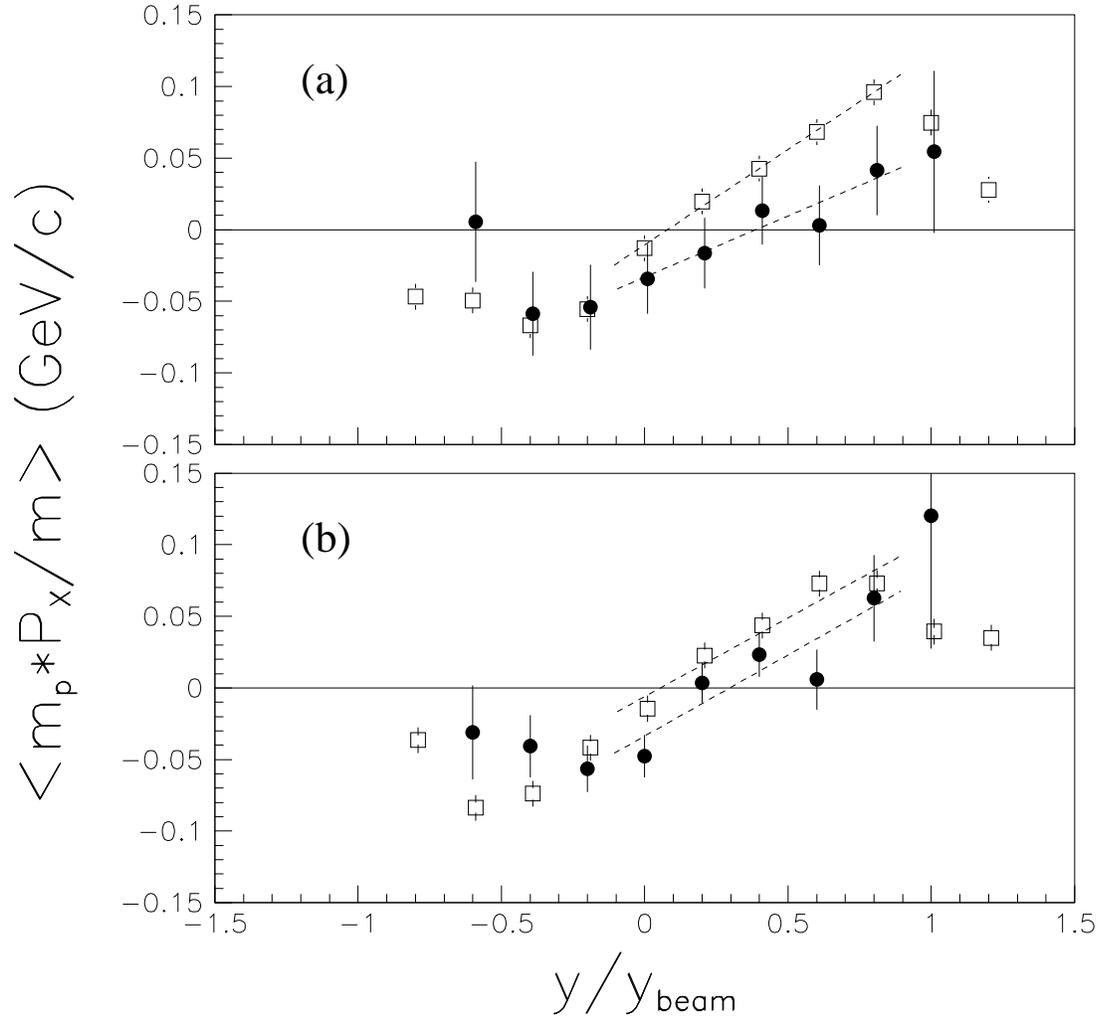,height=6in}}
 \caption { Average in-plane momentum versus normalized rapidity for
(a) data and (b) filtered ARC. Open squares represent protons, closed
circles represent $\Lambda$'s.}
 \label{FLOW}
\end{figure}

The mid-rapidity (-0.25 $\leq y \leq$ 0.25)
$\Lambda$ $m_{T}$ spectrum
is shown in Fig.~\ref{MTLAM}.
In this representation, a 
purely thermal distribution would result in 
a straight line. 
The data of Fig.~\ref{MTLAM}, however, have
the characteristic shoulder indicative
of a collective outward expansion depleting 
the low $m_{T}$ region \cite{sie79,barz91,jeong94,hsi94,lisa95,wang96}.
Nearly twenty years ago, Siemens and Rasmussen
noted that the spectra of protons and pions measured
at $90^{\circ}$ in the center of mass
at the Bevalac showed a pronounced  shoulder at low $p_{T}$, 
which they attributed to 
collective expansion \cite{sie79}. 
By assuming a spherically symmetric thermalized
source at mid-rapidity expanding with
a constant velocity, $\beta_{r}$, they were
able to derive an analytical expression which,
in terms of transverse mass, is:\
\begin{equation}
\frac{1}{m_{T}^{2}}\frac{dn}{dm_{T}} \propto  
e^{-\gamma m_{T}/T}\left[\frac{\sinh\alpha}{\alpha}\left(\gamma
+ \frac{T}{m_{T}}\right) - \frac{T}{m_{T}}\cosh\alpha\right] \,\, ,
\label{RFLOW}
\end{equation}
where $\alpha = \beta_{r}\gamma p/T$ and $\gamma = (1-\beta_{r}^{2})^{1/2}$.

Straightforward application of the
Siemens-Rasmussen expression to the
data of Fig.~\ref{MTLAM} 
gives  $\beta_{r} = 0.36 \pm .02$ and 
T = $51 \pm 1$ MeV with a  $\chi^{2}/\nu$ of 1.1. 
The interpretation of the fitted
parameters is not so straightforward, however. 
Studies of EOS Au+Au data have shown 
that different assumptions about the 
expansion velocity profile are
equally consistent with the observed 
$m_{T}$ spectra.  
In general, the dependence upon
the mean radial flow
velocity and temperature cannot
be expressed in simple analytical
form as above but must be
determined from  Monte Carlo
simulations. In the case of
the present data no attempt has been
made to extract  $\beta_{r}$'s
or temperatures for other velocity 
profiles.  The fit results should 
therefore be viewed as providing 
a quantitative measure of
collective transverse expansion
only within the context of the
velocity profile assumption 
underlying Eq.~\ref{RFLOW}. 

The $m_{T}$ spectrum of
mid-rapidity protons from the same event sample
also has a shoulder. 
An independent fit of the protons results
in a $\beta_{r}$ of $0.42 \pm .01$,
in reasonable agreement with the $\beta_{r}$
for the $\Lambda$'s, with a  $\chi^{2}/\nu$ of
0.75.  The proton's ``temperature'',
however, is much higher: 81 $\pm 1$ MeV.
This difference is consistent with
the picture of cold $\Lambda$'s  
produced just above threshold
undergoing an insufficient number of
rescatterings to come into 
thermal equilibrium with the 
hotter protons.

A two parameter fit of the $\Lambda$'s
obtained by fixing $\beta_{r}$ to zero 
gives T = $106 \pm 5$ MeV
with a $\chi^{2}/\nu$ of 1.9.
Fitting the protons with a pure exponential
results in an inverse 
slope parameter T = $142 \pm 1$ MeV
with a   $\chi^{2}/\nu$ of 1.4.
The rapidity spectrum obtained from
the fit parameters of Eq.~\ref{RFLOW},
assuming a spherically
symmetric source, is indicated by the 
dashed curve of Fig.~\ref{YLAM}.
The result is practically indistinguishable
from a purely thermal 
distribution at the temperature of
106 MeV shown as the solid curve.
The ARC $\Lambda$'s in Fig.~\ref{MTLAM} show no
evidence of a depletion at low $m_{T}$ and
are well fit by simple exponentials. The resulting
inverse slope parameter is 
T = $91 \pm 2$ MeV.
The $m_{T}$ spectra for ARC protons also show no evidence for 
transverse flow. Thermal
fits give T = $121 \pm 1$ MeV  for
filtered and T = $126 \pm 1$ MeV
for unfiltered ARC protons
with  $\chi^{2}/\nu$'s of 1.3 and 1.2,
respectively.

Detection of a majority of the charged particles in the
TPC, along with the presence of directed flow for protons
and heavier fragments, allows for the correlation of
$\Lambda$ production with the event reaction plane.
The standard transverse momentum analysis of Danielewicz
and Odyniec \cite{dan85} has been performed on the
1797 $\Lambda$ events. Details of the analysis 
along with preliminary results obtained from a 
smaller and less clean event sample are given in
Ref.~\cite{qm95}. Briefly, a reaction plane 
for each event is determined 
from protons and nuclear fragments
with $Z\leq8$. The $\Lambda$ transverse momentum is
then projected onto the estimated reaction plane 
and averaged as a function of rapidity over
all events.

The resulting  $\langle p_{x}\rangle$
distributions, 
normalized to the proton mass,
are shown in Fig.~5a 
for $\Lambda$'s (filled circles)
and protons (open squares). 
Both have been corrected for
the $35^{\circ}$  dispersion 
in the estimated reaction plane,
as per the prescription of Ref.~\cite{dan85}.
The conclusion of Ref.~\cite{qm95} has
not changed --- the
$\Lambda$'s ``flow'' in the same direction as the
protons. The dashed lines of Fig.~5 represent
straight line fits to the data over
the region: $0 \leq y \leq 0.8$. 
The slopes, $(m_{p}/m)\times d\langle p_{x}\rangle/dy$,
at mid-rapidity are given in the first
column of Table~1.
The fact that the $\Lambda$ slope is 
consistent with zero within two 
standard deviations is somewhat 
misleading. Backward 
rapidity points have been excluded 
from the fits  because 
reduced acceptance for
$y_{cm} < 0$ leads to systematically lower
values of \mbox{$|\!\langle p_{x}\rangle\!|$}.
This effect can be clearly seen in the
proton data points, which
should be nearly antisymmetric about   
\mbox{$y/y_{beam} = 0$} from symmetry
considerations. The efficiency factors used
to correct the $y$ and $m_{T}$ spectra 
have not  been applied in the case of 
the event-by-event directed flow analysis. 
However, the direction of the correction
can only be toward larger values of
\mbox{$|\!\langle p_{x}\rangle\!|$}, which
bolsters the conclusion of a 
non-zero positive $\Lambda$ flow.

The analysis has also
been performed on the ARC events. 
The results, both before and 
after the GEANT filter,
are given in  Table~1. 
Within statistical uncertainties 
the cascade is in agreement with the data
for both protons and $\Lambda$'s.
A comparison of columns two and
three of the table shows 
(again with large statistical errors)
that the experimental acceptance and 
resolution do not distort 
the extracted flow values.
 
Also shown in Table~1 are flow values extracted
from the relativistic transport code, RVUU,
by the authors of Ref.~\cite{li96}.
Unlike ARC, where flow effects are
produced by preservation of 
two-body scattering planes and an unequal
weighting of attractive and repulsive
orbits \cite{kah94}, RVUU incorporates
a mean field potential for lambdas 
which increases the sidewards
flow in the direction of the protons.
The error bar on the RVUU 
value in Table~1 spans the 
range of flow values that were obtained
in Ref.~\cite{li96} from a
$\pm20\%$ adjustment in the relative
strength of the vector part of
the mean field. 
From the table it is seen that the
two models produce approximately
the same amount of flow for 
protons while ARC $\Lambda$'s
have, perhaps, slightly more
flow. 
Our findings are consistent 
with the  prediction of Ref.~\cite{li96} that the 
magnitude of $\Lambda$ directed flow is 
smaller than proton flow.
Of course, the hypothesis of $\Lambda$ 
flow and proton flow being equal 
cannot be confidently ruled out with 
the present statistics.

The roughly twofold difference 
between the proton $\langle p_{x}/m \rangle$ 
of the present work and the results of 
Ref.~\cite{fopi95} is dominated by the $p_{T}/m > 0.5$ 
selection used in Ref.~\cite{fopi95}.  
For the $\Lambda$'s, both the transverse 
and sidewards flow analyses
are very sensitive to the inclusion 
of combinatoric background. The protons from $\Lambda$ 
decays have momentum vectors
close to their parent particle's due to the low
pion mass. Since non-decay protons, which are
dominant in the formation of the 
combinatoric background,
have both types of flow, 
false $\Lambda$'s also show flow.
In order to obtain  pure 
$\Lambda$ flow it becomes necessary to either: 1)
reduce the background to an insignificant
level as was done in the present analysis;  
or 2) perform a careful analysis of the flow of
the background.
Choosing the first option means 
sacrificing some statistics but
avoids the inevitable complications 
and uncertainties of method 2.
The $\Lambda$ invariant mass plot of 
Ref.~\cite{fopi95} includes a large 
combinatoric background.
It is unclear whether or how much 
their reported $\Lambda$ $\langle p_{x}/m \rangle$ 
are shifted towards their proton $\langle p_{x}/m \rangle$
by this background.

\section{Conclusions}

In summary, we have obtained a high quality sample
of $\Lambda$'s produced in
\mbox{2 $A\,$GeV Ni + Cu collisions.}
To within $15\%$ the
acceptance is well understood, allowing 
efficiency corrections to be made with
confidence.  The $\Lambda + \Sigma^{0}$
production cross section has a linear dependence
upon impact parameter
and the $\Lambda/\pi^{-}$ ratio
is relatively flat versus impact parameter.
A shoulder at low $p_{T}$ in the mid-rapidity
$\Lambda$ tranverse mass spectrum 
is most likely due to collective transverse
expansion. Correlation of the $\Lambda$ momentum
with the reaction plane on an event-by-event
basis reveals a positive directed flow 
consistent with  the predictions of mean field 
theory. The ARC model reproduces the shape
of the $\Lambda$ rapidity spectrum and generates 
sidewards flow in agreement with the data but 
fails to reproduce the shoulders in the 
$\Lambda$ and proton $m_{T}$ spectra and
the dependence of the $\Lambda$ yield upon
event centrality.

\ack{The authors would like to thank David Kahana for providing
the ARC events.
This work is supported in part by the US Department of Energy under 
contracts/grants DE-AC03-76SF00098, DE-FG02-89ER40531, DE-FG02-88ER40408, 
DE-FG02-88ER40412, DE-FG05-88ER40437, and by the US National Science 
Foundation under grant PHY-9123301.}

\vspace*{1in}

\begin{table}[h]
\centering
\label{table1}
\caption {Flow values normalized to the proton mass in units of MeV/c.}
\begin{tabular}{cr@{ $\pm$ }lr@{ $\pm$ }lr@{ $\pm$ }lr@{ $\pm$}l}
\multicolumn{1}{c}{} &
\multicolumn{2}{c}{data}  &
\multicolumn{2}{c}{unfiltered ARC}  &
\multicolumn{2}{c}{filtered ARC}  &
\multicolumn{2}{c}{RVUU} \\ \hline
p & 133  & 10  &  152  & 4 & 141 & 9  & \multicolumn{2}{c}{140} \\
$\Lambda$ & 85 & 43  &  163 & 14 & 113 & 31  &   96 & 17 \\
\end{tabular}
\end{table}

\newpage

\end{document}